\documentstyle[psfig]{mn}

\newif\ifAMStwofonts

\ifoldfss
  \ifCUPmtlplainloaded \else
    \NewTextAlphabet{textbfit} {cmbxti10} {}
    \NewTextAlphabet{textbfss} {cmssbx10} {}
    \NewMathAlphabet{mathbfit} {cmbxti10} {} 
    \NewMathAlphabet{mathbfss} {cmssbx10} {} 
  \fi
  \ifAMStwofonts
    \ifCUPmtlplainloaded \else
      \NewSymbolFont{upmath} {eurm10}
      \NewSymbolFont{AMSa} {msam10}
      \NewMathSymbol{\upi}     {0}{upmath}{19}
      \NewMathSymbol{\umu}     {0}{upmath}{16}
      \NewMathSymbol{\upartial}{0}{upmath}{40}
      \NewMathSymbol{\leqslant}{3}{AMSa}{36}
      \NewMathSymbol{\geqslant}{3}{AMSa}{3E}

      \let\leq=\leqslant \let\le=\leqslant
       \let\ge=\geqslant
    \fi
  \fi
\fi 

\ifnfssone
  \newmathalphabet{\mathit}
  \addtoversion{normal}{\mathit}{cmr}{m}{it}
  \addtoversion{bold}{\mathit}{cmr}{bx}{it}
  \newmathalphabet{\mathbfit} 
  \addtoversion{normal}{\mathbfit}{cmr}{bx}{it}
  \addtoversion{bold}{\mathbfit}{cmr}{bx}{it}
  \newmathalphabet{\mathbfss} 
  \addtoversion{normal}{\mathbfss}{cmss}{bx}{n}
  \addtoversion{bold}{\mathbfss}{cmss}{bx}{n}
  \ifAMStwofonts
    \ifCUPmtlplainloaded \else
      \UseAMStwoboldmath
      \makeatletter
      \new@mathgroup\upmath@group
      \define@mathgroup\mv@normal\upmath@group{eur}{m}{n}
      \define@mathgroup\mv@bold\upmath@group{eur}{b}{n}
      \edef\UPM{\hexnumber\upmath@group}
      \new@mathgroup\amsa@group
      \define@mathgroup\mv@normal\amsa@group{msa}{m}{n}
      \define@mathgroup\mv@bold\amsa@group{msa}{m}{n}
      \edef\AMSa{\hexnumber\amsa@group}
      \makeatother
      \mathchardef\upi="0\UPM19
      \mathchardef\umu="0\UPM16
      \mathchardef\upartial="0\UPM40
      \mathchardef\leqslant="3\AMSa36
      \mathchardef\geqslant="3\AMSa3E

      \let\leq=\leqslant \let\le=\leqslant
       \let\ge=\geqslant
    \fi
  \fi
\fi 

\ifnfsstwo
  \DeclareMathAlphabet{\mathbfit}{OT1}{cmr}{bx}{it}
  \SetMathAlphabet\mathbfit{bold}{OT1}{cmr}{bx}{it}
  \DeclareMathAlphabet{\mathbfss}{OT1}{cmss}{bx}{n}
  \SetMathAlphabet\mathbfss{bold}{OT1}{cmss}{bx}{n}
  \ifAMStwofonts
    \ifCUPmtlplainloaded \else
      \DeclareSymbolFont{UPM}{U}{eur}{m}{n}
      \SetSymbolFont{UPM}{bold}{U}{eur}{b}{n}
      \DeclareSymbolFont{AMSa}{U}{msa}{m}{n}
      \DeclareMathSymbol{\upi}{0}{UPM}{"19}
      \DeclareMathSymbol{\umu}{0}{UPM}{"16}
      \DeclareMathSymbol{\upartial}{0}{UPM}{"40}
      \DeclareMathSymbol{\leqslant}{3}{AMSa}{"36}
      \DeclareMathSymbol{\geqslant}{3}{AMSa}{"3E}

      \let\leq=\leqslant \let\le=\leqslant
       \let\ge=\geqslant
    \fi
  \fi
\fi 

\ifCUPmtlplainloaded \else
  \ifAMStwofonts \else 
    \def\upi{\pi}
    \def\umu{\mu}
    \def\upartial{\partial}
  \fi
\fi

\newcommand{\yray}{$\gamma$-ray}
\newcommand{\yrays}{$\gamma$-rays}
\newcommand{\degree}{\mbox{$^{\circ}$}}
\newcommand{\msol}{$M_{\odot}$}
\newcommand{\HI}{H\,{\sc i}}
\newcommand{\HII}{H\,{\sc ii}}
\newcommand{\kms}{km~s$^{-1}$}
\newcommand{\beam}{beam$^{-1}$}

\def\gapp{\ifmmode\stackrel{>}{_{\sim}}\else$\stackrel{>}{_{\sim}}$\fi}
\def\lapp{\ifmmode\stackrel{<}{_{\sim}}\else$\stackrel{<}{_{\sim}}$\fi}

\title{A search for the radio counterpart of the unidentified \yray\ source 3EG J1410$-$6147}
\author[Doherty et al.]{ M.~Doherty,$^1$
S.~Johnston,$^{1}$ A.J. Green,$^{1}$ M.S.E. Roberts,$^{2}$ R.W. Romani,$^{3}$ 
\newauthor B.M. Gaensler$^{4}$ and F. Crawford$^{5}$  \\
$^1$School of Physics, University of Sydney, NSW 2006, Australia\\
$^2$Department of Physics, McGill University, Montreal, Quebec H3A2T8, Canada\\
$^3$Department of Physics, Stanford University, Stanford, CA 94025, USA \\ 
$^4$Harvard-Smithsonian Center for Astrophysics, Cambridge, MA 02138, USA\\
$^5$Department of Physics, Haverford College, Haverford, PA 19041, USA  
}

\date{\today}

\pagerange{\pageref{firstpage}--\pageref{lastpage}}
\pubyear{2002}

\begin{document}

\maketitle

\label{firstpage}


\begin{abstract}
We have made radio continuum, \HI\ and X-ray observations 
in the direction of the unidentified EGRET source 3EG J1410$-$6147, using the Australia Telescope Compact Array and the Chandra X-ray Observatory.
The observations encompass the supernova remnant (SNR) G312.4--0.4
and the two young pulsars PSRs J1412--6145 and J1413--6141.

We derive a lower distance limit of 6 kpc to the SNR, although interpretation
of positive velocity features in the \HI\ spectrum may imply
the SNR is more distant than 14 kpc.
PSR J1412--6145, with an age of 50 kyr, is the pulsar most likely 
associated with SNR G312.4--0.4. X-rays are not detected from either pulsar and diffuse X-ray emission near the bright western edge of the SNR is weak. Although there is circumstantial evidence that this western region is a pulsar wind nebula (PWN), the embedded pulsar PSR J1412--6145 is apparently not sufficiently powerful to explain the radio enhancement. The origin of the electron acceleration in this region and of the \yrays\ remain unidentified, unless the distance to PSR J1413--6141 is at least a factor of 3 lower than its dispersion measure distance. 

\end{abstract}



\section{Introduction}
Since the discovery of the first unresolved \yray\ source some 30 years
ago, a large number of high-energy sources 
remain without obvious counterparts at other wavelengths.
In the 3rd EGRET catalogue of \yray\ 
sources \cite{hbb+99} there are $\sim$170 sources which are as yet 
unidentified. There have been a variety of approaches used in attempting to 
determine the nature of these unidentified sources. However, due to very 
large positional uncertainties for EGRET sources (of order 1\degree), 
identifying counterparts at other wavelengths is problematic,
particularly in the Galactic Plane where the surface density of sources is 
high. The most powerful technique for identifying sources at other 
wavelengths is a detailed multiwavelength study of individual 
sources (e.g. Roberts et al. 2001)\nocite{rrk+01}. 

There are two \yray\ sources in the southern Galactic Plane near longitude
312\degr. Observations of 2EGS J1418$-$6049 at radio and X-ray
wavelengths have shown that it is likely associated with the
young radio and X-ray pulsar PSR J1420--6048 \cite{rrj01}.
A hard X-ray source embedded in a radio nebula, possibly
a pulsar wind nebula (PWN) may also contribute to 
the \yray\ flux \cite{rrjg99}.
The second \yray\ source, 3EG J1410$-$6147 (2EGS J1412$-$6211),
was tentatively linked with the
supernova remnant (SNR) G312.4$-$0.4 as the most 
plausible radio counterpart \cite{hbb+99}.  This
association was discussed by
Case \& Bhattacharya (1999)\nocite{cb99}
but they did not reach any definitive conclusions, merely speculating that the \yray\ emission could 
be due to either emission from a PWN within the SNR or the acceleration 
of cosmic rays by the supernova shock.
\begin{table*}
\centering
\begin{tabular}{llllrlcrc}
\multicolumn{1}{c}{Name} & \multicolumn{1}{c}{RA} & \multicolumn{1}{c}{Dec} &
\multicolumn{1}{c}{$P$} & \multicolumn{1}{c}{$\dot{P}$} &
Age & \multicolumn{1}{c}{DM} & \multicolumn{1}{c}{d} & \multicolumn{1}{c}{$\dot{E}$}\\ 
 & \multicolumn{1}{c}{J2000} & \multicolumn{1}{c}{J2000} & &
($\times10^{-15}$)& & & &($\times$10$^{34}$) \\
 & \multicolumn{1}{c}{(hms)} & \multicolumn{1}{c}{(dms)} &
   \multicolumn{1}{c}{(s)} & (s s$^{-1}$) & (kyr) &(pc cm$^{-3}$)&
   (kpc) & (erg s$^{-1}$)\\
\hline
PSR J1407$-$6153 & 14 07 57 & $-$61 53 59 & 0.7016 &   8.85 & 1250   & 645 & 9.7 & 0.12 \\
PSR J1412$-$6145 & 14 12 08 & $-$61 45 29 & 0.3152 &  98.65 &   50.6 & 515 & 7.8 & 12.4\\
PSR J1413$-$6141 & 14 13 10 & $-$61 41 13 & 0.2856 & 333.44 &   13.6 & 677 & 10.1 & 56.5 \\
\hline
\end{tabular}
\caption{\label{psrtab}Properties of three radio pulsars in the error box
of 3EG J1410$-$6147}
\end{table*}

SNR G312.4$-$0.4 was discovered by Caswell \& Barnes (1985)\nocite{cb85}
who estimated a lower distance limit of 3.8 kpc 
from ${\rm H}_{2}{\rm CO}$ absorption measurements.
SNR G312.4$-$0.4 appears in the Molonglo Observatory Synthesis 
Telescope (MOST) SNR catalogue \cite{wg96}
and in a detailed study of this region with the MOST at 
843 MHz by Whiteoak, Cram \& Large (1994)\nocite{wcl94}.
At this frequency, the 
SNR has a horseshoe-like morphology, with weaker emission in the 
south. There is a faint arc of emission 
about 12\arcmin\ to the south of the remnant, presumably a blow-out
of part of the shell.
Observations have also been made at 
4.5 and 8.55 GHz with the Parkes radio telescope of both total 
 and linearly polarised intensity \cite{whi93}.
The eastern boundary and the western enhancement of the SNR were
found to be strongly polarised at 8.55 GHz ($\sim$10$-$20\%), while 
lower levels of polarisation were found at 4.5 GHz ($\sim$2$-$5\%). 
This is potentially indicative of a PWN.
Figure 1 shows the MOST image of the SNR overlaid with the 
68\%, 95\% and 99\% confidence
contours for the position of 3EG J1410$-$6147.

There are three pulsars within the 99\% confidence limits
for the position of 3EG J1410$-$6147, all recently
discovered by the Parkes Multibeam Survey (Manchester et al. 2001 and http://www.atnf.csiro.au/research/pulsar/catalogue)\nocite{mlc+01}.
Table \ref{psrtab} lists their
position, period ($P$), period derivative ($\dot{P}$),
characteristic age (given by $P/2\dot{P}$),
dispersion measure (DM), distance (d) determined from the DM using
the electron density model of Cordes \& Lazio (2002)\nocite{cl02},
and spin-down energy ($\dot{E}$).
The DM determined distances are known to be problematic in this part of the
Galactic plane \cite{jkww01} and are likely to be uncertain by a factor of 2.
The two young pulsars both lie (in projection at least) within the SNR shell
and are potential counterparts for 3EG J1410$-$6147.

Given the presence of young, energetic pulsars, the SNR and a putative
PWN, we undertook radio and X-ray observations of this region.
The radio observations were primarily aimed at determining an \HI\
kinematic distance to the SNR. A known distance then enables
us to consider the energetics of the system. In X-rays we hoped to
detect the young pulsars and hence unravel the
nature of the SNR and its PWN.

\begin{figure}
\vspace{10cm}
\caption{\label{egretolay}MOST image of SNR G312.4-0.4 at 843 MHz (Whiteoak \& Green 1996),
overlaid with EGRET confidence contours at 68\%, 95\% and 99\% (Hartman et al. 1999).}
\end{figure}


\section{Observations}
\subsection{Radio Observations}
Radio observations of SNR G312.4$-$0.4 were made with the Australia 
Telescope Compact Array (ATCA; Frater, Brooks \& Whiteoak 1992)\nocite{fbw92}.
The ATCA is an east-west synthesis telescope located near Narrabri, NSW 
and consists of six 22~m antennas on a 6~km track. Five of the antennas 
are moveable along the track, enabling observations to be carried out in 
configurations of different baselines between 31~m and 6~km. The ATCA 
can observe at two different frequencies simultaneously and is 
capable of recording all 4 Stokes parameters at each frequency.

Our observations consist of two separate pointings and were obtained over 
6 observing sessions, with four different array configurations, 
6D, 1.5A, 750C and 375. Details are given 
in Table~\ref{obs-dates}. Each observation was for 12 hours.
Antenna 1 was not functioning during the observations on
2001 Jan 9; hence, only 10 baselines were recorded.

Our observations comprise continuum data split into 32 channels, each of 4 MHz,
for a total bandwidth of 128 MHz centred at 1384 MHz, and \HI\ data 
with a 4 MHz bandwidth and 1024 channels, centred at 1420 MHz. 
All Stokes parameters were recorded for the continuum data, enabling 
polarisation parameters to be calculated. The \HI\ observations were made only in 
total intensity. 

\subsection{X-ray Observations}
We obtained a 9.7 ks exposure of G312.4-0.4 with the Chandra X-ray 
Observatory on 2001 July 26, with pointing centre RA (J2000) 14 12 6.52, Dec (J2000) $-$61 45 44.8.
The four ACIS-I chips, together with two chips from the ACIS-S
array provide significant coverage of the SNR (see Fig.~\ref{xrays}).
The data were reduced
with standard procedures from the CIAO 2.2.1 package. 
No large background flares were observed during this exposure,
so the full time interval was incorporated into the final maps.

\subsection{Other Observations}
We have used several other data sets in this study:
 
\begin{itemize}
\item Radio data from the MOST at 843 MHz \cite{gcly99}.

\item Single dish continuum and \HI\ data for this region, at 21 cm, which
were taken as part of the Southern Galactic Plane Survey \cite{mgd+01}.

\item Images from the Cohen \& Green (2001)\nocite{cg01} study of this
region of the Galactic Plane in the mid-infrared using observations
made with the Midcourse Space Experiment (MSX).
\end{itemize}

\begin{table*}
\centering
\begin{tabular}{llllcc}
Date & Array & \multicolumn{2}{c}{Pointing centre (J2000)} & $\nu_1$ & $\nu_2$ \\
& config. & RA & Dec & (MHz) & (MHz)\\
\hline 
1998 Nov 1 & 6D & 14:13:22 & $-$61:33:31 & 1384 &  \\ 
1999 Nov 27 & 375 & 14:13:22 & $-$61:33:31 & 1384 &   \\
1999 Dec 17 & 1.5A & 14:13:22 & $-$61:33:31 & 1384 &  \\
2001 Jan 9 & 750C & 14:12:00 & $-$61:47:30 & 1384 & 1420\\ 
2001 Jan 12 & 750C & 14:12:00 & $-$61:47:30 & 1384 & 1420\\
2001 Feb 25 & 375 & 14:12:00 & $-$61:47:30 & 1384 & 1420\\
\hline
\end{tabular}
\caption{\label{obs-dates}ATCA Observations of SNR G312.4$-$0.4.  }
\end{table*}


\section{Radio Data Reduction}
\begin{figure*}
\vspace{10cm}
\caption{\label{1384}ATCA greyscale images of SNR G312.4$-$0.4 at 1384 MHz. The images have a resolution of 25\arcsec. The greyscale 
range of image (a) is 0 to 20 mJy \beam\ and that of image (b) is 0 to 
30 mJy \beam. The FWHM of the restoring beam is shown in the lower 
corner of each image. The positions of PSRs J1413$-$6141 and J1412$-$6145 are marked with white 
crosses and seven unresolved sources have been labelled as objects of 
interest as possible counterparts to the EGRET source. The western 
region of the SNR, part of which may be a PWN, has been enlarged to show more detail.}
\end{figure*}
Reduction and analysis of the radio data were carried out with the MIRIAD package using 
standard techniques described in \emph{The Miriad User's Guide} \cite{sk98}.
The primary calibrator PKS B1934$-$638 was used for flux 
density and bandpass calibration.
The secondary calibrator B1329$-$665 was used to solve for antenna 
gains, phases and polarisation leakage terms. After calibration was 
completed, the continuum data was recorded as 13 channels each with 8 MHz bandwidth. The spectral line data have 512 channels, each with a velocity
resolution of 1.6~\kms.

\subsection{Continuum Image}
The continuum data was reduced using mosaicing techniques in order to account for 
the two different pointing centres. The image was formed using a weighting scheme intermediate between natural 
weighting (which minimises noise) and uniform weighting (which minimises 
sidelobe levels). This was achieved by setting the \texttt{robust} 
parameter to 0.5 in the \texttt{invert} task, giving nearly the same 
sensitivity as natural weighting but with significantly lower sidelobes.
The image was deconvolved using a maximum entropy algorithm which is 
implemented in the task \texttt{mosmem} \cite{ssb96}.
The final image has 3\arcsec\ pixels and a beamsize of 25\arcsec.
The rms noise in the image is 0.5 mJy beam$^{-1}$, which is higher
than the theoretical value but not unexpected considering the
excess background emission in the Galactic Plane. This image is shown in Figure~2.

A second image was also formed, combining the ATCA interferometric data with 
single dish data from the Parkes radio telescope \cite{mgd+01}.
There are two possible methods for doing this. One is a non-linear technique 
whereby the data are combined in the $uv$ plane and then jointly deconvolved. 
An alternate method, which is used here, combines the data in the Fourier 
domain after both have been separately deconvolved (e.g. McClure-Griffiths
et al. 2001\nocite{mgd+01}). This was implemented by the 
task \texttt{immerge} in MIRIAD. Including the single dish 
data adds the smoothly varying component of the source and the
diffuse Galactic emission.

Finally, a high resolution image was produced, with 6\arcsec\ 
resolution, using only the baselines with the 6~km antenna.
This resolves out most of the extended emission and allows
better sensitivity for point sources hidden beneath the extended structure.
The rms noise in this image is 0.07 mJy \beam.

\subsection{Polarisation Images}
Stokes $Q$, $U$ and $V$ images were formed by inverting and cleaning each
of the 13 frequency channels separately, using the task \texttt{pmosmem}.  A linear polarisation image, 
$L$, was formed for each channel from the final $Q$ and $U$ images, where 
$L = \sqrt{Q^2 + U^2}$. The position angle image, PA, was
also calculated, where PA = $\frac{1}{2}{\rm tan}^{-1} (U/Q)$.
The $L$ images were then summed over all channels to produce
a final linear polarization image and a bias was removed
to ensure the noise in the image had zero mean.
The final rms noise of this image is 0.14 mJy \beam, significantly better than
for the total intensity.
The resulting polarisation image was divided by the total intensity 
image and clipped where the total intensity emission was less than 
5$\sigma$ (2.5 mJy beam$^{-1}$), to form a fractional polarisation image. This is shown in Figure~\ref{fracpol}.

\subsection{Spectral Index}
In order to compute spectral indices using data from only two wavelengths, 
it is essential that the flux densities be accurately determined. 
For an interferometer, the integrated flux density measured 
depends on the length of the shortest baseline used, in other words, 
the largest spatial scale that is detected. It is, therefore, important that the data contain the same spatial scales 
(e.g. Gaensler et al. 1999; Katz-stone et al. 2000) \nocite{gbm+99,kkld00}.

To match the ATCA and MOST data, we created a new continuum image at 20 cm, using only the pointing 
centred at RA (J2000) 14:12:00, Dec (J2000) $-$61:47:30.
These data contain baselines $\le$~750 m and
have a similar beam ($\sim$~40\arcsec) to the MOST image at 843 MHz.
The two images still
have different $uv$ coverage, so spatial filtering 
was carried out to equalize the spatial content of the two images. 
Extended structure was excluded from the
843 MHz image and the 1384 MHz 
image was smoothed to the lower resolution of the 843 MHz image.

We used two methods for calculating spectral indices, 
spectral tomography and temperature-temperature (T-T) plots and the results were compared, for consistency. Spectral tomography produces maps of the following quantity:  
\begin{equation}
I_t(\alpha_t) = I_{1} - \left(\frac{\nu_{1}}{\nu_{2}}\right)^{\alpha_t}I_{2}
\end{equation}
where $\alpha_t$ is the spectral index, as 
defined by $S_{\nu}\propto \nu^{\alpha}$, and $I_{1}$ and $I_{2}$ are the 
intensities of a given pixel at $\nu_1$ (843 MHz) 
and $\nu_2$ (1384 MHz) respectively \cite{kkld00}. A series of maps was 
made for $-1.0 \leq \alpha_t \leq 0.1$ in steps of 0.1. Any regions of the 
image which vanish into the background have $I_t(\alpha_t) = 0$
and a spectral index of $\alpha_t$. 

T-T plots traditionally involve plotting the brightness temperature of two 
images against each other, pixel-by-pixel. However, the flux density in 
units of Jy beam$^{-1}$ can be used providing the two images have the same 
beam size \cite{boc97}. This method is independent of any zero error, and the slope of the line of best fit is given by
\begin{equation}
\label{tt}
m = \left(\frac{\nu_{1}}{\nu_{2}}\right)^\alpha
\end{equation}
where $\alpha$ is the spectral index. The uncertainty in 
the linear fit may be underestimated as adjacent pixels are not 
independent and there are typically many pixels per beam. This 
effect can be corrected for by rebinning each image, taking 
every $i$th pixel where there are $i$ pixels per beam in each 
dimension \cite{gbm+99}. T-T plots were made for subsets of the 
images at the northern rim of the SNR and also at the brightest part of the western region. A linear least-squares-fit was then 
used to determine the slope for each region. The results were 
checked for self consistency by varying the box size for the plots 
(i.e. the number of pixels) and by shifting the box by a 
few pixels. The slope stayed the same for each respective region and the fit improved with the inclusion of more pixels.  

\subsection{Spectral-line}
After flagging and calibration of the two \HI\ datasets,
we performed continuum subtraction (task \texttt{uvlin})
by fitting a second order polynomial to the line-free channels.
A `dirty' cube was produced using the task \texttt{invert},
selecting only those baselines which were longer than 125 m and
using uniform weighting.
The output cube consists of 150 channels each 2~\kms\ wide
starting at $-120$~\kms. The cube was not cleaned
but was convolved with a gaussian restoring beam of size 45 \arcsec.

We made a continuum image in an identical fashion
using the two datasets and the same parameters in \texttt{invert}.
In this case we cleaned the image appropriately and restored it with
the same gaussian beam of FWHM 45 \arcsec.

The rms noise as a function of channel number was computed using the routine \texttt{imstat}.
The sensitivity varies over the bandpass depending on the temperature
of the \HI\ emission in a given channel. Although the presence of
variable sky structure in the cube will adversely affect the computed rms noise, this was the best way to assess the significance of any 
detected absorption feature.
The data were then
scaled by the total flux in the 
continuum image. This allows the fractional absorption depth
to be computed and the significance of the feature estimated.

An emission profile was obtained using the Parkes single dish data,
for a location just outside the SNR (McClure-Griffiths et al. 2001).

\section{Results}
\subsection{Continuum Image}
Radio continuum images of SNR G312.4$-$0.4 with a resolution of
25\arcsec\ and an rms noise of 0.5 mJy beam$^{-1}$
are shown in Figure~\ref{1384}.
The SNR is seen as an incomplete shell, with a horseshoe-like appearance.
Its brightest regions are the extended western component and the north rim.
The arc-shaped blow-out feature which is visible in the MOST image to the 
south of the SNR can only be seen very faintly here, because of partial primary beam correction. 
The brightest part of the western region of the SNR has a flux density 
of $\sim$30 mJy beam$^{-1}$. It extends over a region of 14\arcmin\
by 7\arcmin\ and has an integrated flux density of $\sim$5 Jy, with the contribution from the shell subtracted. 
The morphology of the western bright region consists of a complicated 
filamentary structure, which is seen for the first time. 
The brightest parts of the northern rim of the SNR have 
flux densities $\sim$20 mJy beam$^{-1}$.

There are several bright, unresolved sources in the image. Seven of the sources (labelled in Figure 2) are now discussed. Source 1 (G312.11--0.2) is non-thermal with a flux
density of 600 mJy at 408 MHz \cite{cb85} and $166\pm3$ mJy at 1384 MHz.
It is classified as extragalactic by Cohen \& Green (2001).
Source 2a (G312.36--0.04) has a flux density of $173\pm6$ mJy at 1384 MHz and is
classified as a compact \HII\ region \cite{cg01}
but the nature of source 2b (flux density $75\pm2$ mJy at 1384 MHz) is unclear although 
it is probably non-thermal.
The flux densities at 408 MHz and 5 GHz given by Caswell \& Barnes (1985) 
are affected by blending between the two sources. Sources 3 to 6 have 1384 MHz flux densities of
$74\pm6$, $59\pm6$, $35\pm7$ and $17\pm8$ mJy respectively. These sources lie within the SNR shell and confusion with the underlying SNR emission contributes to the large error bars.
Source 5 is seen in the infrared and although not classified by
Cohen \& Green (2001)\nocite{cg01} is thought to be a compact \HII\ region.
Source 6 has an X-ray and an infra-red counterpart and is thus likely to be stellar.

The two pulsars listed in Table~\ref{psrtab} are not 
visible in Figure~\ref{1384} but their positions are shown as crosses.
They are detected in the high resolution image and our positions agree,
within the uncertainties, with those from the Parkes Multibeam 
Survey \cite{mlc+01}.
We measured flux densities at 1384 MHz of $0.79\pm0.07$ mJy and $0.73\pm0.07$ mJy for PSRs J1413$-$6141 and 
J1412$-$6145 respectively.  

The long filaments to the north-west of the SNR have a flux
density of $\sim$20 mJy beam$^{-1}$ and are also visible in
the infrared (see section 4.5).

There are several weak background sources in the image which are
almost certainly extragalactic. These are unlikely to be
responsible for the \yray\ emission as their flux density 
is a factor $\gapp 100$ weaker than the known EGRET blazars \cite{msm+97}.
\subsection{Polarisation}
The fractional polarisation image is shown in Figure~\ref{fracpol}.
The image has been masked where the total intensity is less than 
2.5 mJy beam$^{-1}$. The strongest polarisation comes from 
a small area at the southern tip of the western region of the SNR,
which is polarised at around 5$-$10\%. It is offset by about 1\arcmin\ 
from a bright spot in the radio continuum. This is an unusual feature, 
whose origin is unclear.

The absence of any significant polarisation at 1384 MHz is 
somewhat surprising, particularly as the western edge of the SNR
is polarised at the 20\% level at 8.5 GHz \cite{whi93}. 
The lack of polarisation at 1384~MHz can possibly be explained by beam depolarisation, resulting from changes in the degree of foreground Faraday rotation on scales smaller than the synthesised beam. Regions of significant beam depolarisation at this frequency have indeed been determined for nearby regions of the Galactic plane \cite{gdm+01}.

\subsection{Spectral Indices}
T-T plots were made for the western part of the SNR shell and the
northern rim and the results are shown in Figure~\ref{ttplot}. 
As described in Section 3.3, only independent pixels have been used, in 
order to avoid underestimating the uncertainties. 
The two regions have significantly different spectral indices.
The slope of the linear fit for each region is related to the spectral 
index as expressed in equation~\ref{tt}. The slope of the distribution of points 
along the rim gives $\alpha$~=~$-$0.7$\pm$0.1, which is 
typical of synchrotron radiation seen from shell SNRs. The distribution of 
points from the western region gives $\alpha$ = $-$0.19$\pm$0.01.
This value is much flatter than the typical spectral indices for
shell SNRs ($-0.4\lapp \alpha \gapp -0.7$) but lies
well within the typical range expected for a PWN
($0.0\lapp \alpha \gapp -0.3$).

Spectral tomography indicates a spectral index of between $-$0.6 and $-$0.8 
for the northern rim and a value that is much flatter for 
the western region, consistent with the T-T result.
\begin{figure}
\vspace{10cm}
\caption{\label{fracpol}Fractional polarisation for SNR G312.4$-$0.4 at 
1384 MHz. The greyscale ranges from 0\% (white) to 7\% (black). 
Contours show the radio continuum and the levels are 5, 10, 15, 20, 25, 
30, 40 mJy \beam.}
\end{figure}
\begin{figure}
\centerline{\psfig{figure=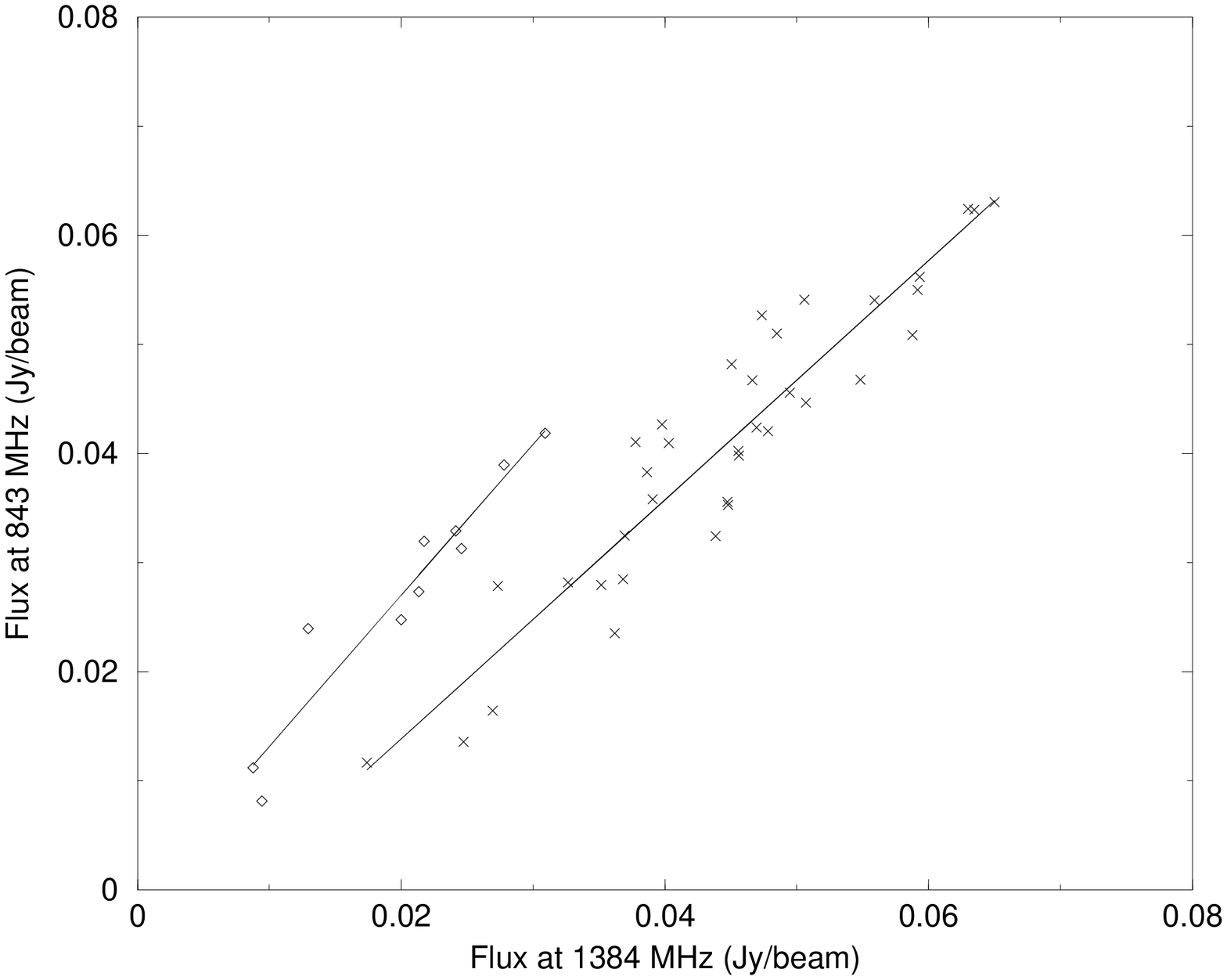,height=6cm,angle=-90}}
\caption{\label{ttplot}T-T plots for two sub-regions of SNR G312.4$-$0.4.
The diamonds show points from the northern rim whereas the crosses show 
points in the western region. The slope of the best fit line to the 
distribution from the rim is 1.4$\pm$0.14, giving $\alpha$ = $-$0.7$\pm$0.1 
and the slope of the best fit line to the points from the western 
region is 1.1$\pm$0.07, giving $\alpha$ = $-$0.19$\pm$0.01.}
\end{figure}

\subsection{X-rays}
\begin{figure}
\vspace{10cm}
\caption{\label{xrays}Smoothed ACIS-I+S2/S3 (2$-$8 keV) image of unresolved emission in the direction of G312.4$-$0.4. The radio contours and pulsar positions (crosses) are shown. Point source subtraction was carried out and the nature of the remaining source is unknown. Only the region with clear structure is significant. The edges of the image are outside the chip coverage.
}
\end{figure}
There are no bright sources in the X-ray image. An application of \texttt{wavdetect}
to the 0.2$-$8 keV data using a threshold SN = 2 detected only
18 sources. Several of these were identified with bright stars; none correlate
with the bright radio point sources. Presumably, most of these sources are coronal
emission from field stars or (for a few hard absorbed sources) emission from
background AGN. Any emission from the G312.4$-$0.4 complex should also be highly absorbed. We therefore formed a 2$-$8 keV image and replaced the counts from the apertures
of all point sources detected in the full-band image with Poisson counts from the local 2$-$8 keV background. The background counts/pixel were generally quite constant across the frontside
illuminated chips. We subtracted the average counts/pixel from chips I0$-$I3 and S2
and a separate background from the back-illuminated S3 chip. No exposure
correction was applied as the fluctuations at the chip boundaries 
obscured any potential diffuse emission. After binning and adaptive smoothing,
the resultant ACIS image is plotted in Figure 5 with the 20 cm radio contours overlaid. A few
regions of excess emission are seen, but at the west limb of the SNR only a
slight excess in the diffuse hard count rate is visible. No significant diffuse emission is seen below 2 keV, indicative of high absorption.

        This diffuse X-ray emission from the western region is marginally detected. In the (unsmoothed) 2$-$10 keV point source-removed image, we find 
105$\pm$24 excess counts in a 6$^\prime \times 6^\prime$ region covering the brightest
portion of the radio image. For typical $\langle n_{HI}/n_e\rangle$ values 10--30, the dispersion measure inferred \HI\ column density to PSR J1412--6145 is comparable to or greater than the total column of Galactic gas in this general direction,
$N_H = 2.2 \times 10^{22} {\rm cm^{-2}}$ \cite{dl90}. Adopting this latter column density and assuming a power law distribution with an index $\Gamma=1.5$, which is generic for the hard 
component of young pulsars and their nebulae, we calculate a flux in X-rays of
${\rm f_X(2-8keV)} = 3.2 \pm 0.7 \times 10^{-13} {\rm erg~cm^{-2}~s^{-1}}$.
This flux value is fairly insensitive to $N_H$ unless it substantially exceeds $10^{23}
{\rm cm^{-2}}$. The ASCA archival images also show some hard X-rays associated with this diffuse source. There is enhanced emission in the range 4$-$10 keV which has a flux of $\sim 4.5 \times 10^{-13}{\rm erg~cm^{-2}~s^{-1}}$. This is consistent with the Chandra results.

We have also looked for unresolved sources corresponding to PSR J1412$-$6145
and PSR J1413$-$6141. No counts are detected at either position.
PSR J1412$-6145$ is close to the optical axis of the ACIS-I and 
the excellent point spread function allows us to put an upper limit on 
the counts (2$-$8 keV), assuming an unabsorbed energy distribution with a power law index $\Gamma=1.5$, which is equivalent to a flux of $6 \times 10^{-15} {\rm erg~cm^{-2}~s^{-1} }$
(3$\sigma$). At the position of PSR J1413$-$6141, the degraded 
point spread function allows an upper limit of
$1.8 \times 10^{-14} {\rm erg~cm^{-2}~s^{-1} }$(3$\sigma$), insensitive to a moderate range of $N_H$ values.

\subsection{Infrared}
\begin{figure}
\vspace{10cm}
\caption{\label{8mn}8 $\mu$m mid-infrared image with 6\arcsec\ pixels,
overlaid with radio contours at 20cm. The greyscale ranges from
0.78 MJy sr$^{-1}$ to 10 MJy sr$^{-1}$. Contour levels are 5, 10, 15, 
20, 30, 50, 90 and 120 mJy \beam.}
\end{figure}

Figure~\ref{8mn} shows radio contours overlaid on the 
mid-infrared 8 $\mu$m image from the MSX satellite \cite{cg01}. 
A very striking feature of Figure~\ref{8mn} is the presence of thermal and radio continuum
filaments running diagonally across the image.
The bright filaments in the infrared are slightly offset from those in 
the radio. Cohen \& Green (2001)\nocite{cg01} have made an extensive
comparison of this region in the radio and the infra-red. They discuss the
nature of the filaments (and other sources) in detail.

Although there is some diffuse infra-red emission throughout the SNR
it is clear that there are no major \HII\ regions within its boundaries.
In particular the western part of the shell, which has a flat spectral
index in the radio, is not an obvious thermal source.
As previously mentioned, the radio sources 5 and 6 have infrared
counterparts and there is also a 
cluster of strong infrared sources close to the top rim of the SNR 
which are likely to be \HII\ regions or stars.

\subsection{\HI\ Spectral Line Data}
\begin{figure}
\centerline{\psfig{figure=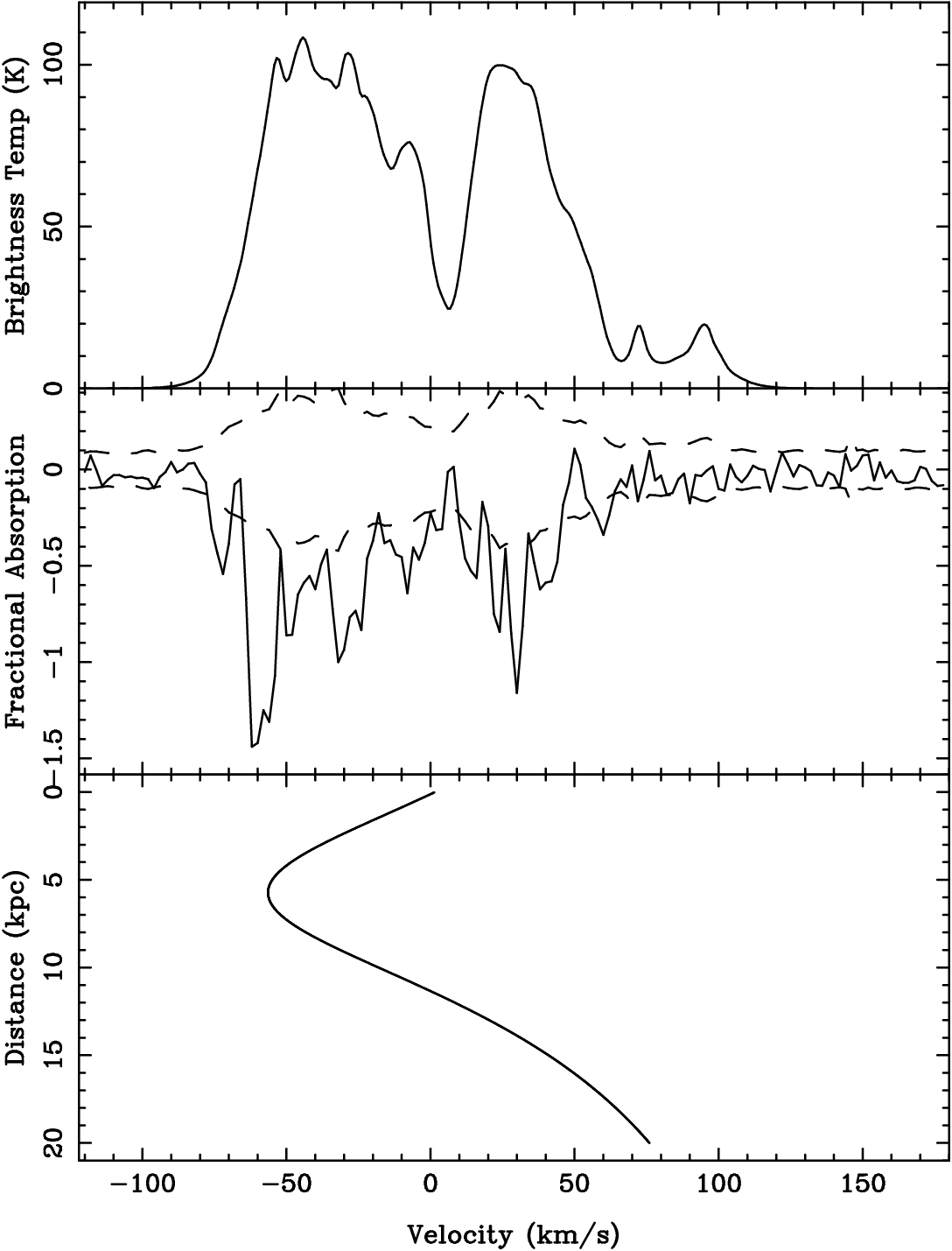,height=10cm}}
\caption{\label{p1}Emission (top) and absorption (middle) spectra in 
the direction of the point source P1. The dashed line in the middle panel shows the $1\sigma$ envelope as a function of velocity channel. The bottom panel shows the
conversion of velocity to distance at this galactic longitude.}
\end{figure}

We compare our resulting absorption spectra with the emission
profile in that direction, obtained from the Parkes data (McClure-Griffiths et al. 2001).
Figure~\ref{p1} shows the emission profile and directly underneath it,
the absorption spectra for Source 1 (as marked in the
continuum image in Figure~\ref{1384}). The brightness temperature is
scaled to match the value in the Kerr et al. (1986)\nocite{kbjk86}
\HI\ emission survey of this part of the Galaxy.
The dashed line is the 1$\sigma$ envelope, plotted as a 
function of velocity channel and clearly shows that the noise in the line is about 4-5 times higher than the rms noise off-line.
Below the spectra we show the rotation curve appropriate
for this longitude. To compute the rotation curve
we use an analytic expression given by the best fit model of the Galactic rotation curve from Fich, Blitz,
\& Stark (1989)\nocite{fbs89}.
We adopt the standard IAU parameters \cite{klb86} for the distance
to the Galactic centre ($R_{0}$ = 8.5~kpc) and the solar orbital velocity
($V_{0}$ = 220~\kms). The rotation curve of the Galaxy is the
azimuthally smoothed average of its velocity field
and thus implicitly assumes circular orbits around the Galactic centre.

\begin{figure}
\centerline{\psfig{figure=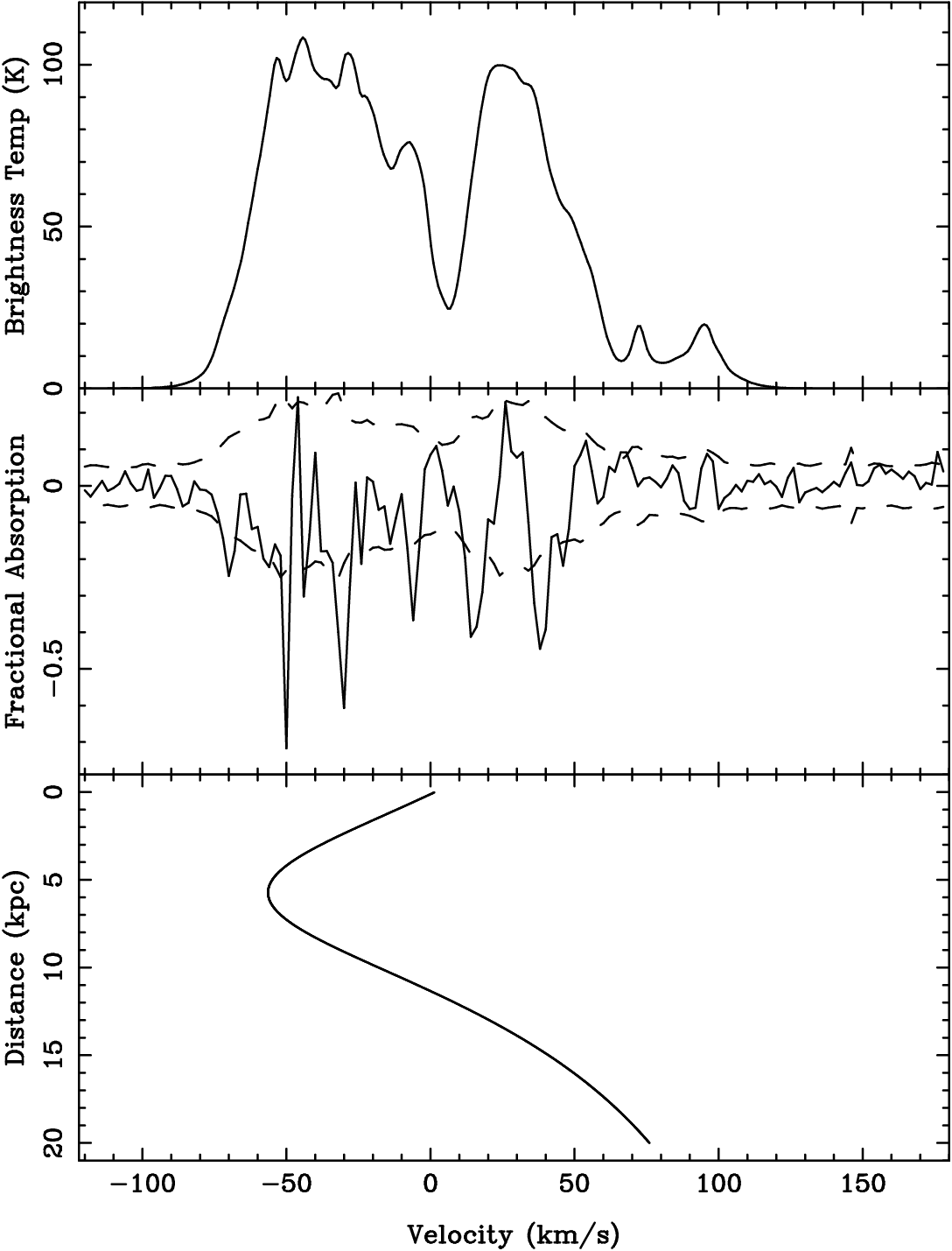,height=10cm}}
\caption{\label{pwn}Emission (top) and absorption (middle) spectra in
the direction of the western part of the SNR. The dashed line in the middle panel shows the $1\sigma$ envelope as a function of velocity channel. The bottom panel shows the
conversion of velocity to distance at this galactic longitude.}
\end{figure}

The emission spectrum is somewhat unusual in that it shows
relatively weak emission at zero velocity. High
brightness temperature gas is seen out to the tangent
point at $-$55~\kms. At positive velocities we see 100~K gas
out to +40~\kms\ and two further weak emission features at
+70 and +95~\kms. One would not expect to see absorption against
these two weak features.

The absorption spectrum for Source 1 (Figure~\ref{p1})
displays deep absorption at all negative velocities out to the tangent point.
Absorption with a similar optical depth to the negative velocity
features is also seen from +20 to +50~\kms.
Source 1 shows absorption against all the main emission
features and it is almost certainly extragalactic, confirming the
classification of Cohen \& Green (2001).
The absorption spectrum for Source 2a is very similar to that of
Source 1 and it seems probable that it is
also extragalactic. This is at odds with the compact \HII\ region classification
of Cohen \& Green (2001) but it is likely that their
infrared to radio flux ratio (on which their classification is based)
is too high because of confusion with
the (foreground) filamentary structure.
None of the other point sources were of sufficient flux density to
show significant absorption in their spectra.

Figure~\ref{pwn} shows the absorption spectrum in the direction of
the western region of the SNR.
The spectrum exhibits deep absorption features at negative velocities out to the tangent point. Features with a smaller optical depth are also seen,
at 20~\kms\ and 40~\kms, with a significance of 4$\sigma$. 
Given the deeper absorption 
features out to the tangent point, we can certainly
impose a lower distance limit of 6 kpc. Any conclusions beyond 
this depend then on the interpretation of the features at 
20~\kms\ and 40~\kms. If these absorption features are real, a 
lower distance limit of 14 kpc is implied. 

The absorption spectrum towards the
northern rim is very noisy and cannot provide any 
direct evidence of whether or not the northern rim and
the western region are physically connected.

\section{Discussion}
Given the uncertainties in the interpretation of the absorption spectrum,
in the following discussions we adopt a distance of 6$d_6$~kpc
to the SNR complex.

\subsection{Origin of the SNR}
As the main shell is quite symmetrical,
we estimate its centre to be at RA (J2000) 14 13 00,
Dec (J2000) $-$61 43 00. The radius of the shell is then 13\arcmin, 
equivalent to a physical radius of 23$d_6$ pc.
Assuming an ambient density of $n \sim 1$ cm$^{-3}$, the mass swept up by the shell is $\sim$1200$d_6^3$ \msol. This implies that the SNR is well into the 
Sedov-Taylor (adiabatic) phase of expansion as it has swept up more than 
20 times the mass ejected \cite{dj96}.

We can relate the initial conditions of the SN to the radius of the 
expanding shock front by
\begin{equation}
R(t) = 5.39\left(\frac{E_{51}t^2}{n}\right)^{0.2}  {\rm pc} 
\end{equation}
where $E_{51}$ is the kinetic energy of the explosion in units of 
10$^{51}$ erg, $t$ is the time after the explosion in kyr
and $n$ is the density of the medium into which the SNR is 
expanding in particles cm$^{-3}$.
Assuming that the age of the SNR is the same as PSR J1412$-$6145 (50 kyr),
we find the ratio $E_{51}/n$~$\sim$~0.6$d_6^5$.
Alternatively, the age of PSR J1413$-$6141 (13 kyr), yields a
ratio of $E_{51}/n$~$\sim$~9$d_6^5$.
Given the uncertainties of the pulsar ages and the SNR distance,
both these values are well within expected bounds for the
adopted distance $d = 6$ kpc. However, if $d \ge 14$ kpc then 
PSR J1413$-$6141 may be ruled out as 
the progenitor star for the SN explosion, as $E_{51}/n$ becomes
unreasonably large.

The pulsar velocities can be 
calculated by assuming, in turn, that each was the progenitor 
star for the SNR. This implies a transverse velocity of 
$v=218 d_6$ \kms\ for PSR J1412$-$6145 and
$v=288 d_6$ \kms\ for PSR J1413$-$6141.
Again, both these values are entirely consistent with pulsar
velocities in general \cite{cc98}, even if $d \ge 14$ kpc.

\subsection{A PWN around PSR J1413$-$6141?}
We might expect to see a radio PWN surrounding PSR J1413$-$6141, a young,
highly energetic pulsar.
Gaensler et al. (2000)\nocite{gsf+00} discuss two types of 
PWN which can form, a static PWN which is confined 
by the gas pressure of the surrounding 
interstellar medium and a bow-shock PWN confined by ram 
pressure from the motion of the 
pulsar through the medium. The condition for a static PWN is given by
\begin{equation}
nV^5 \le 4 \times 10^9 \dot{E}_{34}/t^2
\end{equation} 
where $V$ is the pulsar velocity in \kms, $\dot{E}_{34}$ is the spin-down energy in units of $10^{34} {\rm erg s^{-1}}$ and the other symbols are as defined earlier.
Taking a lower limit on $n$ of 0.003 cm$^{-3}$,
the velocity of the pulsar would need to be less than 210~\kms\ for
this condition to be met. If this is the case, the size of the static
PWN can be computed via
\begin{equation}
\label{static}
R_{\rm static} = 0.14\left(\frac{\dot{E}_{34} t^3}{n}\right)^{0.2} {\rm pc}
\end{equation}
yielding a radius of 4.8 pc.
This PWN would
be resolved in our image for any sensible pulsar distance but
distinguishing the PWN from the remaining SNR structure is virtually
impossible.

If the pulsar velocity is in excess of 210~\kms,
it will have overtaken its static 
PWN resulting in a bow-shock PWN with a characteristic scale size
\begin{equation}
R_{bow-shock} = 0.63\left(\frac{\dot{E}_{34}}{nV^2}\right)^{0.5} {\rm pc}
\end{equation}
Even for a lower limit on $n$ of 0.003 cm$^{-3}$, the bow shock PWN
would be unresolved in our image.
Assuming a typical PWN spectral index $\alpha = -0.3$ and integrating from 10 MHz to 100 GHz, if the PWN is unresolved then the efficiency of the pulsar is
given by
\begin{equation}
\epsilon = 4.8 \times 10^{-6} \frac{d^2 S_{1.4}}{\dot{E}_{34}}
\end{equation}
where $S_{1.4}$ is the PWN flux density at 1384 MHz in mJy.
In our case we measure a flux density of 0.97 mJy at
the pulsar position as compared to a catalogue value of 0.51 mJy for
the pulsed flux.
Generously assuming a PWN flux of 0.5 mJy
the efficiency is then $\le$ 1.5 $\times$ 10$^{-6} d_6^2$.
This is more than an order of magnitude less than expected for pulsars
of this age \cite{gsf+00}.

We therefore conclude that PSR J1413--6141 has a velocity
less than 200~\kms\ and has formed a static PWN whose surface
brightness is confused with other emission or below our detection limit.
This again points to the fact that 
PSR J1413--6141 is likely not to be associated with the SNR. 

\subsection{Evidence for a PWN near PSR J1412$-$6145?}
There is circumstantial evidence that the western part
of the SNR is a PWN. Whiteoak (1993) reports 20\% polarization
at 8 GHz and, as we have derived, the spectral index of this
region is rather flat. However, the lack of polarization at both
5 GHz and 1.4 GHz is puzzling. This either implies a substantial
rotation measure or that the 8 GHz polarization is over-estimated.
 On the other hand, if this \textit{is} a PWN we need to determine
the nature of its power source.

The integrated flux density of the PWN is $\sim$ 5 Jy. 
Using the measured spectral index of $-$0.2, we estimate the integrated
radio luminosity from 100 MHz to 100 GHz to be $\sim$1.1$\times 10^{34}
d_6^2$ erg s $^{-1}$. The PWN covers 
an area of 12 $\times$ 24 $d_6^2$ pc and so
we assume a value of $\sim$9~pc for its radius in the subsequent discussion.
We can relate the radio luminosity, $L_R$, of a PWN to the
spin down energy of the central source via $L_R=\epsilon \dot{E}$
where $\epsilon$ is the efficiency. Gaensler et al. (2000) showed
that in young pulsars $\epsilon$ was of order 10$^{-4}$, and hence
one requires the central pulsar to have $\dot{E} \gapp 10^{38} d_6^2$ erg s$^{-1}$
in our case. A pulsar of this $\dot{E}$ should be highly visible in X-rays. Furthermore, the two known pulsars in the field have $\dot{E} \sim10^{35}$ erg s$^{-1}$, which is clearly not high enough to power a PWN of such size and luminosity. The lack of any obvious high $\dot{E}$ pulsar in the Chandra image, and the low values of $\dot{E}$ for the known pulsars, are inconsistent with the calculated parameters for a PWN. If d $\sim$ 14 kpc, then both $L_R$ and the size of the region
become improbably large for a PWN. The source of the electrons powering this radio enhancement remains unclear.

\subsection{X-ray emission}
The limits on the X-ray luminosity of PSRs J1412--6145 and
PSRs J1413--6141 are 2.6$\times$10$^{31}$ erg s$^{-1}$ and
7.8$\times$10$^{31}$ erg s$^{-1}$ respectively assuming a distance
of 6 kpc. This implies an X-ray efficiency of less than 2.1$\times$10$^{-4}$
and 1.4$\times$10$^{-4}$. These limits are similar to the expected values for these sources from the Possenti et al. (2002)\nocite{pccm02} relation and are also in line with other pulsars of similar ages, eg. the Vela pulsar and the
recently discovered pulsar PSR J2229+6114 \cite{hcgh+01}.

The X-ray luminosity of the diffuse emission near 
PSR J1412--6145 is 1.4$\times$10$^{33}$ erg s$^{-1}$.
This is comparable to the luminosity of the diffuse nebula surrounding PSR B1951+32 in CTB 80, although the $\dot{E}$ of this pulsar is higher than that of PSR J1412--6145 by a factor of $\sim$30 \cite{sof95}. Therefore the weak diffuse X-ray emission seen from this region in the ASCA and CXO data could plausibly be a PSR J1412--6145 powered PWN.

\subsection{\yray\ emission}
The 3rd EGRET catalogue \cite{hbb+99} gives the integrated 
flux of the \yray\ source as \emph{F} = 64.2 $\times$ 10$^{-8}$ 
photons cm$^{-2}$ s$^{-1}$ with a photon index of 2.12.
Using an upper cut-off of 10 GeV, this gives a \yray\ 
luminosity, $L_{\gamma} = 2.1\times 10^{36}$$d_6^{2}$ erg s$^{-1}$,
assuming isotropic radiation or
$1.7\times 10^{35}d_6^{2}$ erg s$^{-1}$ into 1 steradian.
Tompkins (1999) \nocite{tom99} has computed a variability index for the EGRET sources, finding a value of $0.33_{-0.17}^{+0.22}$ for 3EG J1410$-$6147. Nolan et al. (in prep.)\nocite{nol02} using 
a
revised statistic similarly find the value $0.32_{-0.24}^{+0.32}$. These results 
show
only that the source is significantly less variable than the bright EGRET 
blazars, 
suggesting a pulsar or diffuse emission source of $\gamma$-ray flux.

The latter value of $L_{\gamma} = 1.7\times 10^{35}d_6^{2}$ erg s$^{-1}$ is 139\% of the $\dot{E}$ for PSR J1412$-$6145 
and 30\% of the $\dot{E}$ for PSR J1413$-$6141, even at d = 6~kpc.
Generally one observes an increasing efficiency with 
decreasing $\dot{E}$ and
PSR B1055$-$52 is the most efficient \yray\ pulsar with 
efficiency of 15\% \cite{thom98} and $\dot{E}$ a factor of 10
lower than PSR J1413$-$6141. For PSR J1412$-$6145 to be the \yray\ source, a small beaming angle
and a high efficiency would need to be invoked. The efficiency of
PSR J1413$-$6141 is an order of magnitude higher than
expected for its $\dot{E}$ and therefore it is unlikely that either pulsar powers the \yray\ source unless the distance is substantially smaller that that inferred from dispersion measure and \HI\ measurements.

With the existence of a PWN uncertain and {\it prima facie} evidence for \yray\ emission from the pulsars unconvincing,
the origin of the \yray\ source 3EG J1410$-$6147 remains a mystery.
It is tempting to speculate on the existence of a third young pulsar in this region. It would have $\dot{E} \sim 10^{37} {\rm erg s^{-1}}$, sufficient to explain the \yrays\ and power the radio PWN. The main problem with this explanation is that its X-ray efficiency would have to be extremely low ($\le 10^{-6}$) to explain its non-detection in our X-ray observations. It would then then have the lowest value, by an order of magnitude, of $L_x/L_{\gamma}$ of any of the known pulsars.
A plausible alternative model could be based on the interaction of the western edge of the SNR with a nearby molecular cloud (eg. Aharonian, Drury \& Volk (1994) and Esposito et al. (1996)).\nocite{adv94} \nocite{ehks96} 


\section{Conclusions}
We have made a continuum image of the SNR G312.4--0.4 and
surrounding regions at 1384 MHz, with a resolution of 25\arcsec.
We also made observations in the 21 cm \HI\ line and produced 
absorption spectra to obtain a lower distance limit to the SNR of 6 kpc.
If absorption features at positive velocities are real, the SNR is
located beyond 14 kpc.
Although the bright, western portion of the SNR resembles a PWN, at these distances PSR J1412--6145 seems inadequate to power the observed enhancement to the radio flux.
The lack of a radio PWN 
around the 13 kyr PSR J1413$-$6141 indicates it is 
likely to be slow moving and hence not formed with SNR  G312.4--0.4.
We suggest that the 50 kyr pulsar, PSR J1412$-$6145,
originated in the same explosion as the SNR.

The two young pulsars superimposed on the SNR are not detected in the X-rays, although the implied 10$^{-4}$ limits on the X-ray efficiencies are not strongly constraining.
Neither of these two pulsars can account for the \yray\ luminosity
unless abnormally high efficiency, small beaming angle 
and a low distance are invoked.

The most likely interpretation of the multi-wavelength data of this region is
(i) the SNR is large, old ($\sim$ 50,000 yrs) and distant, with 
PSR J1412$-$6145 as the likely progenitor,
(ii) the bright western region of the SNR is most probably 
an amorphous, flat spectrum part of the SN shell and
(iii) the \yrays\ are unrelated to the SNR.

The \yray\ source 3EG J1410$-$6147 still
remains unidentified.
The launch of the Gamma-ray Large Area Space Telecope in 2005 
will reduce the size of the positional error box to 1\arcmin\ and provide
a resolution to this puzzle.

\section*{Acknowledgements}
The Australia Telescope is funded by the Commonwealth of 
Australia for operation as a National Facility managed by the CSIRO.
We thank Naomi McClure-Griffiths for making the data from 
the SGPS available. The MOST is owned and operated by the University of Sydney, supported by the ARC and the University of Sydney.
This research has made use of data obtained from the High Energy Astrophysics Science Archive Research Centre (HEASARC) provided by NASA's Goddard Space Flight Centre. Support for this work was also provided by CXO grant G01-2070X.

\bibliography{modrefs,psrrefs}
\bibliographystyle{mn}

\label{lastpage}
\end{document}